\documentclass[a4paper,fleqn,usenatbib]{mnras}

\usepackage{newtxtext,newtxmath}

\usepackage[T1]{fontenc}
\usepackage{ae,aecompl}

\usepackage{amssymb} 
\usepackage{amsmath} 
\usepackage{hyperref}
\usepackage{booktabs,longtable}
\usepackage{graphicx} 
\usepackage{wasysym} 
\usepackage{lscape} 
\usepackage{caption} 
\usepackage{wrapfig,lipsum,booktabs} 
\usepackage{enumitem} 
\usepackage{mathtools} 
\usepackage{overpic} 
\usepackage{graphbox} 
\usepackage{tabularx} 
\usepackage{caption} 
\usepackage[table]{xcolor} 
\usepackage{overpic} 
\usepackage{pict2e} 
\usepackage{caption} 
\usepackage{tikz} 
\usepackage{transparent} 

\newcommand*\corser{core-S{\'e}rsic}
\newcommand{\Sersic}{S\'{e}rsic}
\newcommand*\Msun{$M_{\odot}$}

\definecolor{fancygray}{gray}{0.3}
\definecolor{connector}{RGB}{203,117,176}

\newlist{cutenumerate}{enumerate}{1}
\setlist[cutenumerate,1]{
  label={\arabic*},
  leftmargin=*,
  align=left,
  labelsep=0.05cm,
}

\title
[The missing core mass -- fine structure relation] 
{
 Connecting traces of galaxy evolution: the missing core mass -- morphological fine structure relation
} 

\author[P. Bonfini et al.]
{
 \parbox{\textwidth}{
  P. Bonfini$^{1}$\thanks{E-mail: \texttt{p.bonfini@crya.unam.mx}},
  T. Bitsakis$^{1}$,
  A. Zezas$^{2,3,4}$,
  P.-A. Duc$^{5}$,
  E. Iodice$^{6}$,
  O. Gonz{\'a}lez-Mart{\'i}n$^{1}$,
  G. Bruzual$^{1}$,
  and
  A. J. Gonz{\'a}lez Sanoja$^{1}$\thanks{Visiting summer student} 
 }
 \vspace{0.4cm}\\
 \parbox{\textwidth}{
  $^{1}$Instituto de Radioastronom{\'i}a y Astrof{\'i}sica,
  Universidad Nacional Aut{\'o}noma de M{\'e}xico, Morelia 58089,
  M{\'e}xico.\\
  $^{2}$University of Crete, Physics Department \& Institute of Theoretical \& Computational Physics, 71003 Heraklion, Crete, Greece\\
  $^{3}$Foundation for Research and Technology-Hellas, 71110 Heraklion, Crete, Greece\\
  $^{4}$Center for Astrophysics, 60 Garden Street, Cambridge, MA 02138\\
  $^{5}$Observatoire Astronomique de Strasbourg, Universite de Strasbourg, CNRS, UMR 7550, 11 rue de l'Universit{\'e}, F-67000 Strasbourg, France\\
  $^{6}$INAF-Astronomical Observatory of Capodimonte, via Moiariello 16, Naples, I-80131, Italy
 }
}

\date{MNRAS --- Accepted 2017 October 12.}

\pubyear{2017}

\begin{document}
\label{firstpage}
\pagerange{\pageref{firstpage}--\pageref{lastpage}}
\maketitle

\begin{abstract}
 Deep exposure imaging of early-type galaxies (ETGs) are revealing the second-order
 complexity of these objects, which have been long considered uniform,
 dispersion-supported spheroidals.
 ``Fine structure'' features (e.g.\ ripples, plumes, tidal tails, rings) as well as
 depleted stellar cores (i.e.\ central light deficits)
 characterize a number of massive ETG galaxies, and can be interpreted as the result
 of galaxy-galaxy interactions.
 We discuss how the timescale for the evolution of cores and fine structures are
 comparable, and hence it is expected that they develop in parallel after the major
 interaction event which shaped the ETG.
 Using archival data, we compare the ``depleted stellar mass'' (i.e.\ the mass missing
 from the depleted stellar core) against the prominence of the fine structure
 features, and observe that they correlate inversely.
 This result confirms our expectation that, while the Super Massive Black Hole
 (SMBH) binary (constituted by the SMBHs of the merger progenitors) excavates the
 core via three-body interactions, the gravitational potential of the newborn
 galaxy relaxes, and the fine structures fade below detection levels.
 We expect the inverse correlation to hold at least within the first Gyr from the
 merger which created the SMBH binary; after then, the fine structure evolves
 independently.
\end{abstract}



\section{Introduction}
\label{Introduction}

Owing to the development of advanced instrumentation during the last decades, we
are realizing that early-type galaxies (ETGs) deviate from the iconic picture of a
perfectly spheroidal and dynamically relaxed system.
For example, integral field unit surveys such as SAURON
\citep[e.g.][]{cappellari:2007,emsellem:2007} and ATLAS$^{3D}$ \citep{ATLAS3D}
recently showed  that ETGs are not purely ``dispersion supported'' systems (i.e.
their stars only follow disordered orbits), but also present a rotational
component.

Higher-order characteristics of ETGs represent precious information which
potentially allow to  address the different patterns which lead to their formation,
within the widely supported scenario of hierarchical galaxy evolution
\citep[e.g.][and references therein]{white,springel,hopkins,hopkins:2010}, in which
elliptical galaxies are the end-products of violent galaxy interactions.
In this work we concentrate in particular on two attributes of ETGs:
fine structures and depleted stellar cores.

\vspace{-0.2cm}

\begin{center}
\emph{Fine Structures}
\end{center}

\vspace{-0.2cm}

\noindent
Fine structures are extremely faint \citep[24--27 mag/arcsec$^{2}$; e.g.][]{duc:2011}
features which manifest in ETGs as a mixed collection of morphological structures,
including e.g.\ ripples, plumes, tidal tails, ``boxy'' isophotes, or rings.
Such features, originally identified by \cite{schweizer:1988} as a peculiarity
of a few objects, are now recognized to be ubiquitous once an appropriate image depth
is reached \citep[e.g.][]{{duc:2011, spavone:2017}}.
Fine structures depict the remnants of the last interaction which
shaped the observed morphology of the galaxy, and are expected to fade out progressively
as the gravitational potential relaxes
\citep[][]{barnes:1988,hernquist:1988,hernquist:1989}.
They therefore represent a unique tool for studying galactic archaeology, and are
ideal to investigate the history of dry (i.e. relatively gas-free) mergers, where the lack
of star-formation limits the possibility to estimate the time elapsed from the event.

The seminal papers on the subject could be considered the works by
\cite{schweizer:sigma} and \cite{schweizer:1992}, in which the authors provide
a semi-empirical definition of a ``fine structure parameter'' $\Sigma$,
which they visually estimated on [unsharp-masked] images by identifying the
number and/or strength of structures.
Currently,
several projects are creating extended, deep mosaics of nearby galaxies
with the ultimate intent of studying the traces of interactions and
low-brightness galaxy satellites, such as e.g.\ the Next Generation Virgo Cluster
Survey \citep[NGVS;][]{ferrarese:NGVS}, the Fornax Cluster Deep Survey
\citep[FCDS; e.g.][]{iodice:FCDS}, the Mass Assembly of early-Type GaLAxies with
their fine Structures (\textsc{Matlas}; P.I.: P.A.~Duc), or the
VST survey of Early-type Galaxies in the Southern hemisphere
(\textsc{Vegas}; P.I.: E.~Iodice).

\vspace{-0.2cm}

\begin{center}
\emph{Depleted Stellar Cores}
\end{center}

\vspace{-0.2cm}

\noindent
Luminous ($M_{B} < -20.5$ mag) ETGs present one additional
idiosyncrasy
which distinguishes them from the smaller-mass ones: they are characterized
by the presence of a ``core'', i.e.\ a central luminosity
deficit (not due to dust obscuration) with respect to the inward extrapolation
of the outer light profile \citep[e.g.][]{lauer:1985}.

According to the generally accepted scenario, depleted stellar cores are thought
to be related to the dry merger of galaxies.
In this context, the Super Massive Black Holes (SMBHs) at the center of each of
the merging progenitor are dragged by dynamical friction towards the center of the
gravitational potential and eventually form a SMBH binary at the center of
the merger remnant.
The basic idea is that three body interactions between the stars and the SMBH
binary transfer energy to the stars, hence ejecting them from the center of the
newly formed galaxy, creating the central stellar mass deficit
\citep[][]{begelman:1980}.
This hypothesis is supported by the observed scaling relations between the SMBH
mass and the core size or the ``missing stellar mass''
\citep[e.g.][]{lauer:2007a,dullo:2014}.
The energy is transferred to the interacting stars at the expense of the SMBH
binary, whose orbital separation shrinks along the process, leading the binary
to coalescence while the core is excavated.

An important caveat is that the SMBH binary scouring is not expected to happen in
a gas-rich environment.
In fact, in this case the gas would disperse a significant fraction of the
energy of the SMBH binary in lieu of the stars.
Moreover, any excavated core could be replenished by central star formation
\citep[e.g.][]{mihos:cusp,barnes,hopkins:2008,hopkins:2009a}.
This caveat is consistent with the observation that core galaxies are almost exclusively
radio-loud objects \citep[e.g.][]{capetti:2005,capetti:2006,capetti:2007,richings},
given that radio-loud AGNs are long thought to be cold-gas starved systems
in a low activity state \citep[e.g.][]{best:radio}.

In this context, the amount of missing stars --- or ``mass deficit'' --- can be
quantified from the difference between the extrapolation of the outer radial 
profile (indicative of the pristine galaxy) and the observed profile (with an 
excavated core; see panel \emph{C-2} in Figure \ref{figure:M_ratio_Sigma_All}).
The surface brightness profile of the spheroidal component of galaxies hosting a
core can in fact be described by a smooth connection
between an inner power-law and an outer \Sersic{} function 
\citep[the ``\corser{} model'';][]{graham:corser}.

\medskip

\noindent
We have therefore a coherent theoretical framework --- the hierarchical merging
scenario --- able to simultaneously explain: 1) the emergence of fine structure
features in post-merger massive galaxies, 2) the excavation of stellar cores,
and 3) the merging of SMBH binaries at their centers.
What was still lacking, is a coherent \emph{observational} investigation which
would ultimately connect the merger history of massive ETGs with the fine structure,
prominence of core, and central SMBH.

With the present work, we demonstrate that these aspects are indeed intimately
connected.
In $\S$\ref{Timescale Considerations} we discuss how the timescale for the
creation of the core and the disappearance of fine structures are consistent,
in $\S$\ref{Measurements of cores and fine structures} we present observational results
supporting the core -- fine structure connection, in $\S$\ref{Discussion}
we discuss our interpretations, while we in \ref{Conclusions} we summarize our results.

\section{Timescale Considerations}
\label{Timescale Considerations}

\noindent
According to the hierarchical galaxy formation scenario, massive ETGs are
mostly assembled through mergers, whose traces --- the residual fine
structure features --- progressively fade out with time.
At the same time, in the case a dry merger event, a core is excavated by a
coalescing SMBH binary at the center of the newly formed galaxy.
How do the timescale of these two processes compare?

In a recent numerical simulation, \cite{khan:2012a} showed that (at least in
the case of minor mergers of disk galaxies) SMBHs with masses $10^6$--$10^7$
\Msun{} will completely coalesce within 1--2~Gyr after the merger.
This figure encompasses the whole time span from the merger to the actual
SMBH merger.
In a related simulation, the same authors \citep[][]{khan:2012b} showed that,
while the total coalescence time is $\sim$2.9~Gyr, the SMBH scouring phase
lasts $\sim$1~Gyr (after which the SMBH pair rapidly merges via gravitational
wave emission).
In general, there seems to be consensus on that, in dry mergers, the scouring
phase alone can last a few Gyr \citep[for a review, see][]{colpi}.

On the other hand, the average survival time of fine structures is a parameter
much more difficult to estimate; simulation-wise because it requires to reproduce
a great variety of features, and observationally-wise because of the low surface
brightness of the structures compared to the main body of the galaxy.
Despite of the obstacles, the literature regarding simulations aimed at
interpreting the interaction and merger debris is rich.
The variety of these idealized simulations yielded diverse results,
but estimations of survival times in both the spatial-- and the phase--space
agree in that fine structures should remain coherent for at least 1~Gyr
\citep[e.g.][]{hibbard:1995,feldmann,johnston:2001,johnston:2008,michel-dansac,peirani,torrey,duc:simulation}.
Studying color differences between the fine structures and the host galaxy,
\cite{schweizer:1992} estimated the average survival time to 1--2~Gyr.
Using a wider sample and deeper exposures, \cite{duc:2011}
argue that, although an upper limit could be set at 5~Gyr, the
vast majority of the collisional debris would fall back onto the main body of
the galaxy within a couple of Gyr \citep[e.g.][]{hibbard:1995}.
The aforementioned studies additionally show that the specific timescale clearly
depend upon the type of merger which generated a given fine structure.
Structures predominantly associated with ``wet'' (i.e.\ gas rich) events, such as tidal
tails and plumes, are relatively short-lived (1--2~Gyr), while structures potentially
associated with dry mergers such as shells \citep[e.g.][]{paudel:shells_dEs}
could survive up to 4~Gyr.

\medskip

\noindent
Thus, the theoretically estimated timescale for the merger remnants fine structures
to significantly [if not completely] fade out is at least sufficiently large to
complete the formation of cores through SMBH scouring.
In other words, it is expected that fine structure coexists with SMBH binaries
at different stages in their process of excavating the core.
In fact, the two processes should initially advance in parallel: while the SMBH
binary (created soon after the merger event) proceeds with its scouring action,
the fine structure features gradually disappear due to the relaxation of the
gravitational potential.
Since however the fine structure life span is on average larger than the timescale
of core excavation, at some point this co-evolution could cease.
After about $\sim$1 Gyr, while the core will not expand significantly, the galaxy
will still keep reducing its fine structure.

\section{Measurements of cores and fine structures}
\label{Measurements of cores and fine structures}

\begin{table*}

 \begin{tabular*}{\textwidth}{c@{}c@{\hskip 0.5cm}c}

  \begin{overpic}[width=0.25\textwidth, height=0.22\textheight]
   {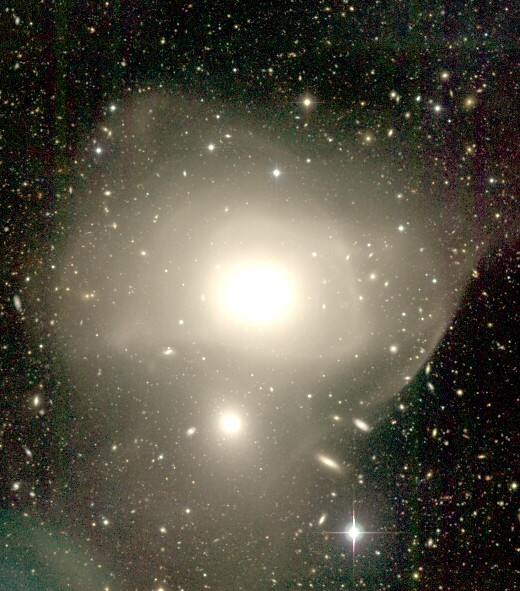}
   \put(55,90){\textcolor{white}{NGC~3640}}
   \put(05,90){\textcolor{white}{A-1}}
  \end{overpic}

  &

  \begin{minipage}{0.43\textwidth}
    \vskip -4.3cm 
    \begin{overpic}[width=\textwidth, height=0.26\textheight]
    {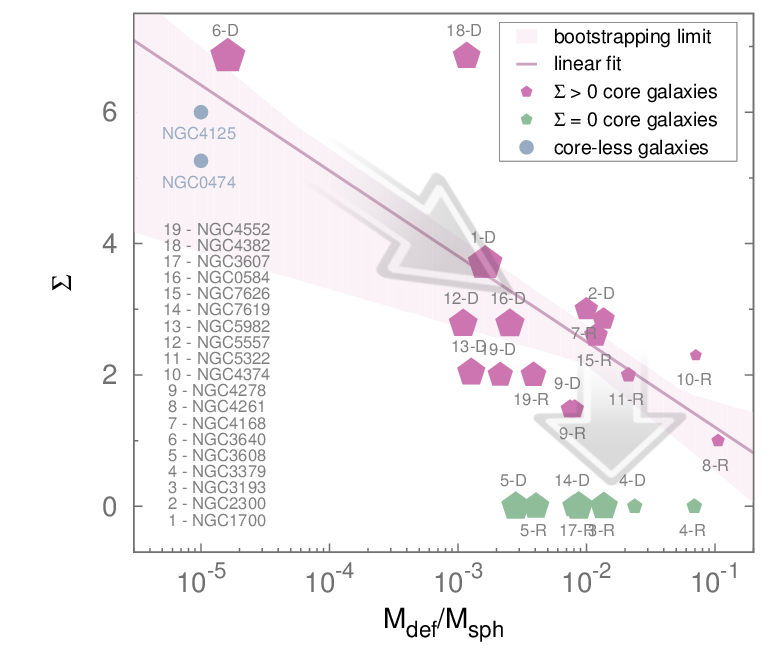}
   \end{overpic}
  \end{minipage}
  
  & 

  \begin{overpic}[width=0.25\textwidth, height=0.22\textheight]
   {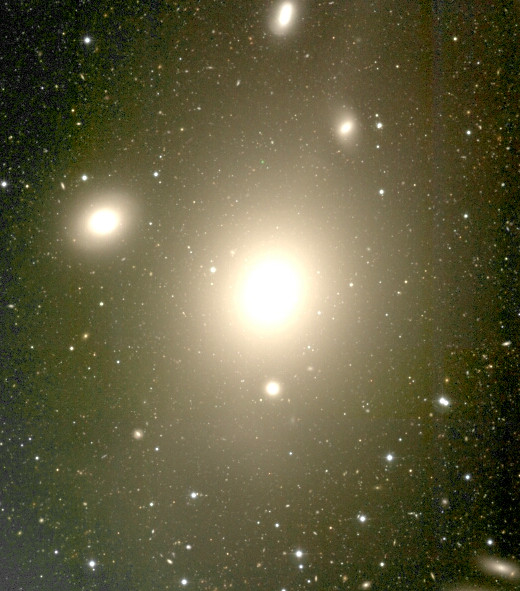}
   \put(55,90){\textcolor{white}{NGC~4261}}
   \put(05,90){\textcolor{white}{C-1}}
  \end{overpic}
 
  \\
  
  \begin{overpic}[width=0.27\textwidth, height=0.252\textheight]
   {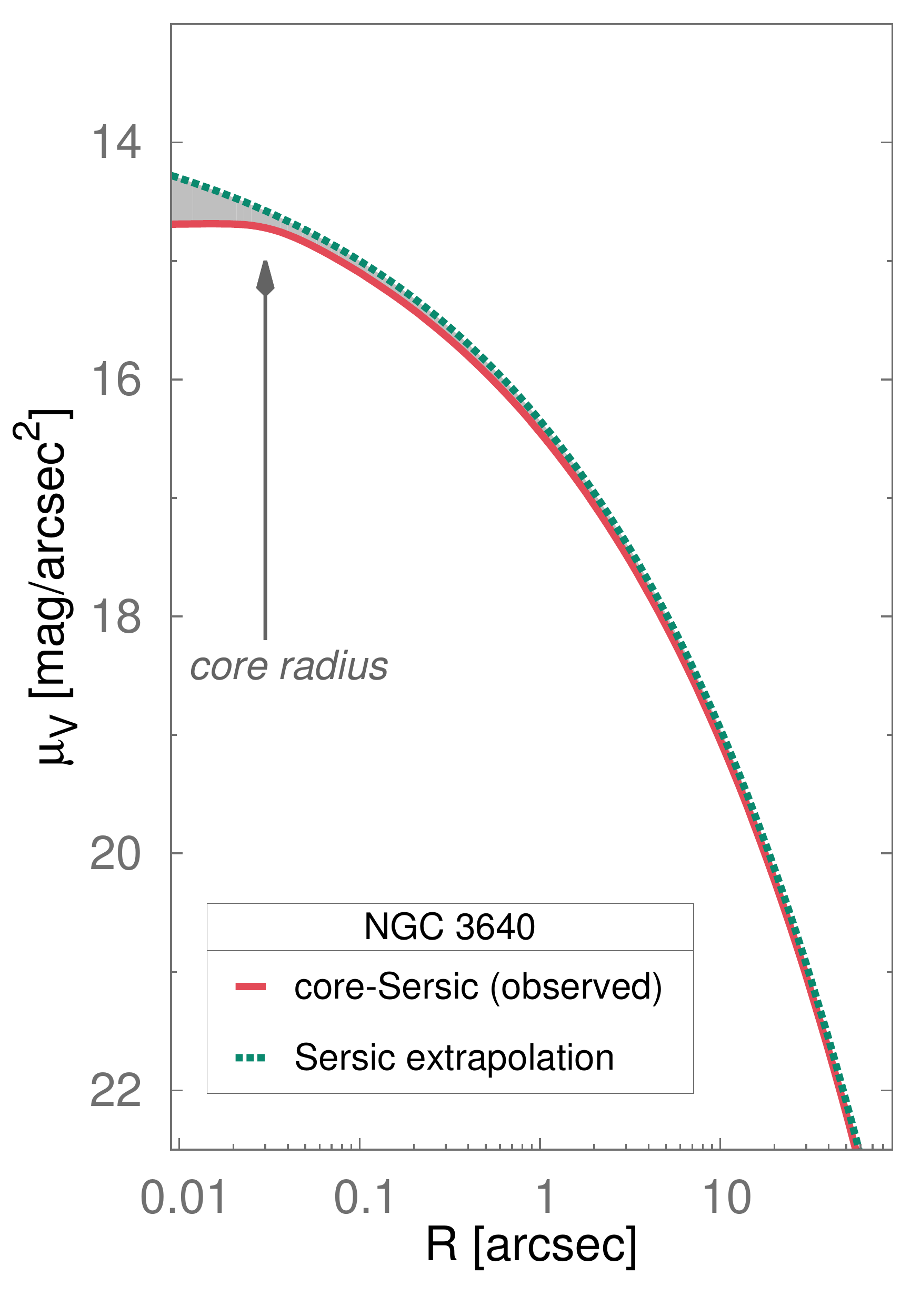}
   \put(20,90){\textcolor{fancygray}{A-2}}
  \end{overpic}
   
  &

  \begin{minipage}{0.43\textwidth}
   \vskip -5.7cm 
   \begin{overpic}[width=\textwidth, height=0.26\textheight]
    {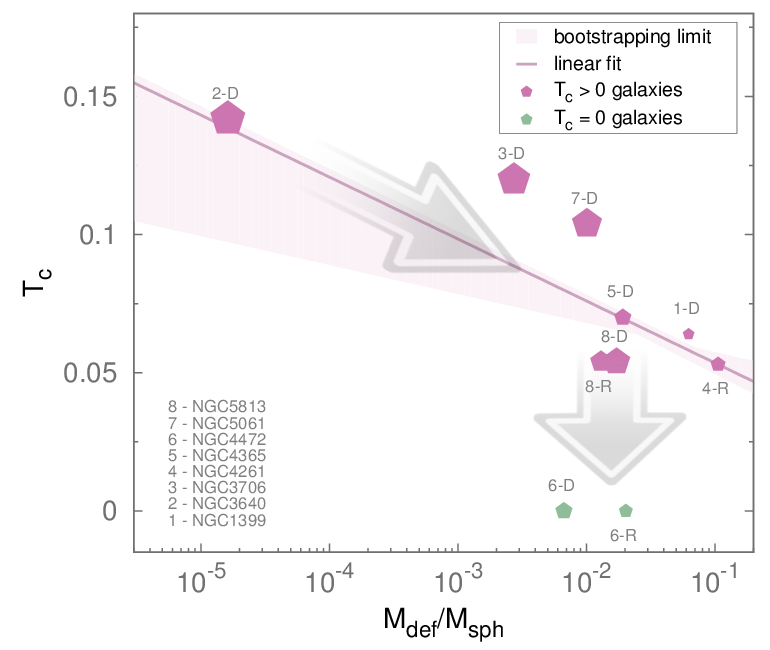}
   \put(19,73){\textcolor{fancygray}{B-2}}
   \end{overpic}
  \end{minipage}

  &
  
  \begin{overpic}[width=0.27\textwidth, height=0.252\textheight]
   {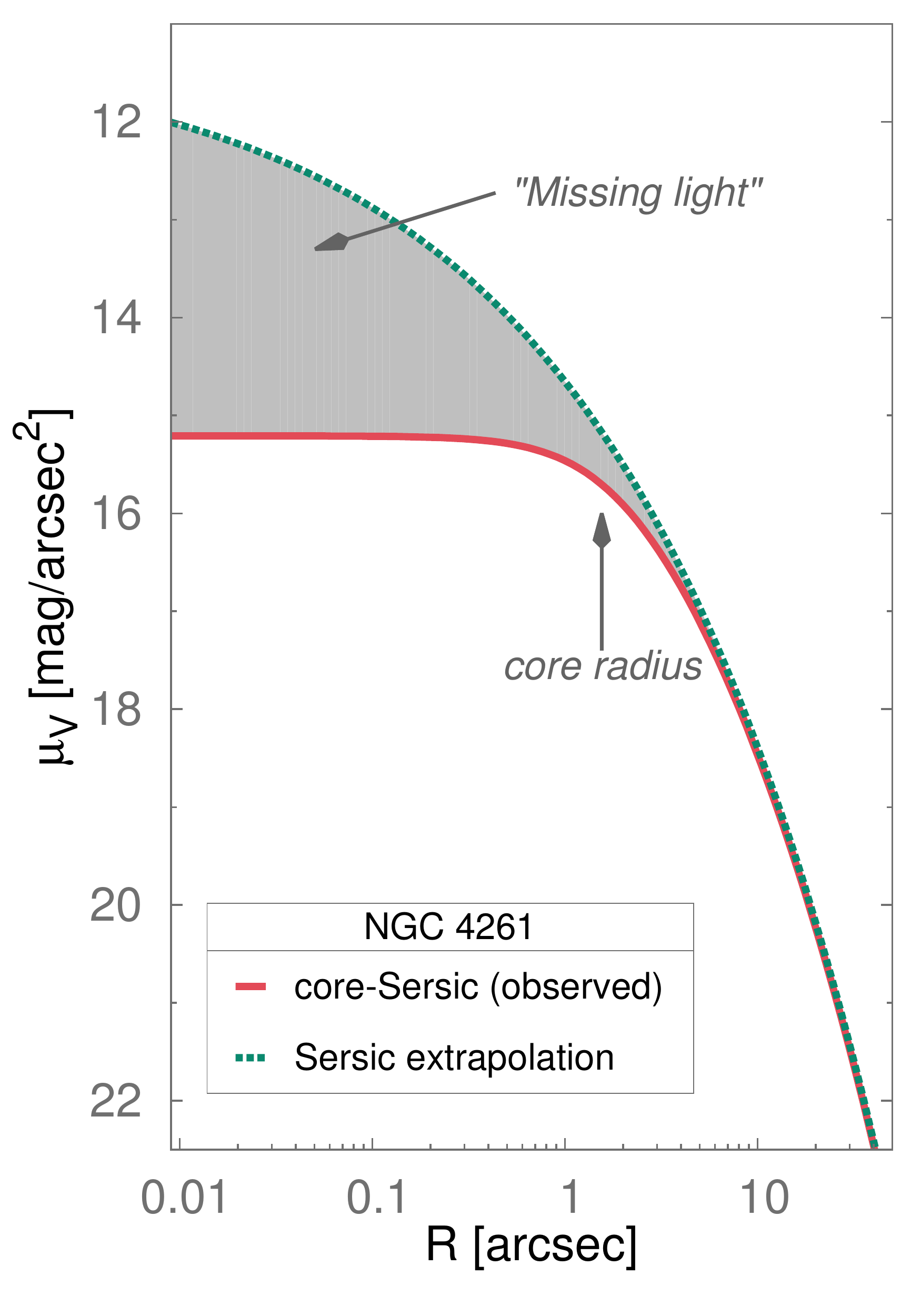}
   \put(20,90){\textcolor{fancygray}{C-2}}
  \end{overpic}
  
  \\

 \begin{tikzpicture}[overlay]
  \node[fill=none, rotate=0.0, minimum width=0.1cm,minimum height=0.1cm] (b) at (10.15,8.45){\textcolor{connector}{
    \Huge{\rule{1.0cm}{1.1pt}\hspace*{-2pt}\raisebox{-4.9pt}{|}}}
  };
  \node[fill=none, rotate=180, minimum width=0.1cm,minimum height=0.1cm] (b) at (3.4,12.15){\textcolor{connector}{
    \Huge{\rule{2.1cm}{1.1pt}\hspace*{-2pt}\raisebox{-4.7pt}{|}}}
  };
 \end{tikzpicture}

 \begin{tikzpicture}[overlay, fill opacity = 0.4]
  \node[fill=none] (box) at (4,12.1) {{\transparent{0.8}\includegraphics[width=0.5cm, height=0.4cm,angle=0.000]{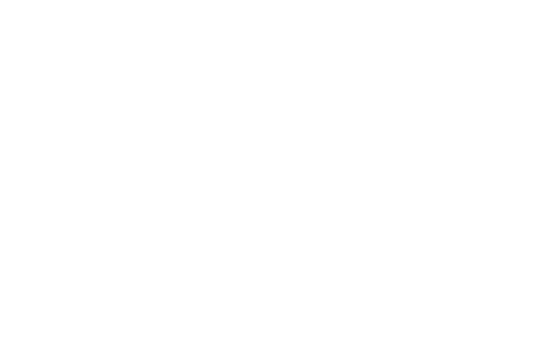}}};
  \node[fill=none, minimum width=0.1cm,minimum height=0.1cm] (b1) at (4,12.1){\textcolor{fancygray}{B-1}};
 \end{tikzpicture}

 \end{tabular*}

 \captionof{figure}[M_ratio_Sigma_All]{
  Relation between fine structure and depleted stellar mass, and suggested evolutionary
  tracks.
  \\
  \emph{Central panels (B) ---}
  Fine structure parameter \citep[$\Sigma$;][\emph{B-1}]{schweizer:sigma}, and
  tidal parameter \citep[T$_{c}$;][\emph{B-2}]{tal} as a function of relative
  mass deficit ($M_{def}$/$M_{sph}$).
  The pink pentagons represent the data points for the core galaxies with detected
  fine structures ($\Sigma > 0$ or T$_{c} > 0$), while the green pentagons represent data
  values for the comparison sample of core galaxies without fine structure
  ($\Sigma = 0$ or T$_{c} = 0$).
  The relative mass deficits were calculated from the \corser{} 
  profiles reported by \cite{richings} and \cite{dullo:2014}, as schematically
  shown in the side panels.
  The size of the data points is inversely proportional to the reliability of the core,
  for which we adopted the ratio between the core size and the seeing FWHM of the relevant
  observation as a proxy.
  The reference for each data point has been indicated with a label ``R'' for
  \cite{richings}, or ``D'' for \cite{dullo:2014}.
  Notice that the two studies have 2 (1) galaxies in common in the $\Sigma > 0$
  (T$_{c} > 0$) sample, and 2 (1) in the $\Sigma = 0$ (T$_{c} = 0$) sample.
  The solid pink line represents the 1-to-1 fit to the $\Sigma > 0$ (T$_{c} > 0$) sample,
  while the light pink shaded area shows the limits of all the possible curves determined
  through a 2-points (1-point) bootstrapped linear fit.
  The dark blue circles in panel \emph{B-1} represent the galaxies classified as
  core-\emph{less} by \cite{richings}, but which could potentially host an unresolved core
  or develop one in the future, as suggested by their $B$-band luminosity
  (see $\S$\ref{Measurements of cores and fine structures}).
  The gray arrows show the proposed evolutionary track, according to which the
  core and the fine structure initially co-evolve (moving diagonally along the
  plot), while --- once the core has been completely excavated --- a galaxy should move
  vertically down as the fine structure gradually disappears.
  \\
  \emph{Side panels (A \& C) ---} \textsc{Matlas} false-color images (\emph{A-1} and \emph{C-1})
  and surface brightness radial profiles (\emph{A-2} and \emph{C-2}) of two galaxies at
  the opposite extremes in terms of both fine structure and core depletion.
  The red solid curves represent the best-fit \corser{} parameters
  from \citet[NGC~3640, left]{dullo:2014} and \citet[NGC~4261, right]{richings}.
  The green dashed curves represent the inward extrapolation of the \Sersic{} part
  of the \corser{} profile.
  We indicate with a vertical arrow the location of the best-fit core radius
  (i.e.\ the transition point between the inner power law and the outer
  \Sersic{}), while the grey shaded areas show how the ``missing light''
  is calculated as the area between the \corser{} curve and its \Sersic{}
  extrapolation.
  \label{figure:M_ratio_Sigma_All}
 }

\end{table*}

\vspace{0.3cm}

\noindent
We defined a sample of nearby ETGs galaxies, for which we could retrieve
literature values for both the fine structure and the stellar mass deficit.
The ``bottleneck'' for this selection was dictated by the scarcity of studies
reporting numerical evaluations of diverse fine structures.
For our purpose, we used the semi-quantitative ``fine structure parameter''
($\Sigma$) defined by \citet[see $\S$\ref{Introduction}]{schweizer:sigma}.
We additionally considered the ``tidal parameter'' ($T_c$) by \cite{tal}, which
however represents a simplified statistics based on the presence of distortions
on the residual images after subtracting a [symmetric] galaxy model,
similarly to the only other studies we are aware of, nominally \cite{vandokkum}
and \cite{mcintosh}.
We stress that, despite of its name, the tidal parameter does \emph{not}
exclusively describe ``tidal tails'', but dishomogeneities in general
(see Equations 1 and 2 of \citealt{tal}).

For the 69 objects in \cite{schweizer:sigma}, we searched in the literature for
\corser{} fits to their surface brightnesses, and found 19 among the samples
of \cite{richings} and \cite{dullo:2014}.
Of those galaxies, 14 display fine structure ($\Sigma$ > 0): these
objects represented the primary sample for the current study (9 have core parameters
from \citealt{dullo:2014},
7 from \citealt{richings}, 2 are in common).
The remaining 5 galaxies did not present any evident structure ($\Sigma$ = 0): we
used this as a comparison sample, keeping in mind that the lack of structure
might be due to the limited sensitivity of the data used in
\citet[][40--60~min exposure images from the 0.9~m KPNO telescope]{schweizer:1992},
rather than an intrinsic feature.
Our small sample is representative of the range of $\Sigma$ in \cite{schweizer:sigma},
and of the depleted mass in typical core galaxies.
Similarly, the sub-sample of 52 galaxies by \cite{tal} with disturbed morphology
($T_c$ > 0) was cross-correlated against \citealt{dullo:2014} and \citealt{richings},
obtaining 6 and 2 matches, respectively (1 in common).
In this case, the comparison sample of featureless objects ($T_c$ = 0) consisted
only in one, common galaxy.

We calculated the spheroid mass ($M_{sph}$; mass associated to the ``bulge'' of
the galaxy, excluding any disk structure) by integrating the \corser{}
profiles by \citealt{dullo:2014} and \citealt{richings}
to obtain the total luminosity, and then converting to stellar mass by
assuming a mass-to-light ratio for a uniformly old
(12~Gyrs), solar metallicity stellar population (as appropriate for massive
early-type galaxies, e.g. \citealt{ETG_age_metallicity})
To derive our mass-to-light ratios we used the ``Worthey model interpolation
engine'' applet\footnote{
  \url{http://astro.wsu.edu/worthey/dial/dial_a_model.html}
}
based on the evolutionary models by \cite{M_L}, where we adopted the default
Salpeter initial mass function prescriptions.
Similarly, the mass deficit ($M_{def}$) was inferred from the difference
between the full \corser{} profile, and the extrapolation
of its \Sersic{} part
\citep[the latter being indicative of the pre-depletion profile;][]{graham:corser}
--- see side panels in Figure \ref{figure:M_ratio_Sigma_All}.

In the central panels of Figure \ref{figure:M_ratio_Sigma_All}, we show the
fine structure parameter (panel \emph{B-1}) and tidal parameter (panel \emph{B-2})
as a function of the mass deficit normalized by the spheroid mass
($M_{def}$/$M_{sph}$).
The scatter in the plots is due to the uncertainty on both quantities:

 \begin{cutenumerate}[label={\arabic*})~]
  \itemsep0.0em  

  \item The parametrisation of fine structure is the main source of uncertainty,
        due both to the lack of a robust definition, and to the dependence
        of $\Sigma$ (or $T_c$) on the image data quality and depth.
        A good example is NGC~5557, for which \cite{schweizer:sigma} reported
        $\Sigma$ = 2.87, while more recent and deeper observations revealed additional
        structure, raising $\Sigma$ to 6.00 \citep{duc:2011}.
        Such depth biases are expected to similarly affect all data points,
        hence maintaining the observed distribution.
        Moreover, one can reasonably assume that any present structure was detected with
        similar efficiency within the \cite{schweizer:sigma} or \cite{tal} samples,
        except possibly for the \mbox{$\Sigma$ = 0} (\mbox{$T_c$ = 0}) galaxies
        (as discussed above).

  \item There is often a significant mismatch in the recovered depleted mass due
        to differences in fitting techniques.
        For the galaxies they have in common (other than the ones shown in
        Figure \ref{figure:M_ratio_Sigma_All}), the mass deficit derived
        from \cite{richings} and \cite{dullo:2014} differ
        on average by $\sim$40\%.
        
 \end{cutenumerate}

\noindent
In the same panels, we show with a solid pink line a linear fit to the $\Sigma$ > 0
and $T_c > 0$ samples, respectively of the form:

\vspace{-0.5cm}
\begin{eqnarray}
 \Sigma = c_\Sigma	+ m_\Sigma	\times log(M_{def}/M_{sph})
 \label{equation:Sigma} 
\end{eqnarray}
\vspace{-0.5cm}

\vspace{-0.5cm}
\begin{eqnarray}
 T_c = c_T + m_T \times log(M_{def}/M_{sph}) 
 \label{equation:T_c} 
\end{eqnarray}
\vspace{-0.5cm}

\noindent
which yielded the best-fit parameters $c_{\Sigma} = -0.10$ and $m_{\Sigma} = -1.30$,
and $c_{T} = 0.03$ and $m_{T} = -0.02$.
These fits have been weighted by the reliability of the core measurement, with the
weights estimated as the ratio between the core size and the seeing FWHM of the
relevant observation.
In order to exclude the possibility that the 2 highest $\Sigma$ galaxies (nominally
NGC~3640 and NGC~4382) drive the relation, we applied a 2-points bootstrapping.
In practise, we devised an iterative procedure in which two random points were
removed from the sample at each iteration and the fit was re-performed, until
all possible combinations were covered.
For $T_c$ we performed a similar test, but adopting a 1-point bootstrapping,
given the smaller number of data.
This analysis yielded a range of best-fit relations which span within the
the shaded pink area in panels \emph{B-1} and \emph{B-2}, representative of the
error on our best-fits.
Additionally, we calculated the Pearson correlation coefficient for the fit of
Equation \ref{equation:Sigma}, obtaining $R = -0.71$ and $-0.44$ for the full
sample and after excluding the 2 points with the largest $\Sigma$, respectively.
For the fit of Equation \ref{equation:T_c}, we obtained $R = -0.85$ and $-0.75$
for the full sample and after excluding the point with the largest $T_c$, respectively.
A test based on the Fisher $z$ transformation showed that, for both fits, 
the Pearson coefficients obtained with the full or the reduced samples are
consistent within the 95\% confidence level.
This indicates that there is no significance difference by excluding
the aforementioned 2 (1) data points, and further confirms the solidity of the
linear relations.

As a further test of our analysis, we wanted to check if galaxies which
have not yet developed a core consistently present high levels of fine structure.
\cite{richings}, apart from the \corser{} galaxies we referred up to now,
also reported 41 objects whose spheroidal component was successfully fit
with a simpler \Sersic{} profile: this implicitly classifies them as core-\emph{less}.
However not all core-\emph{less} galaxies are suitable to our test: we
need to select only those which are indeed ``pre-destined'' to develop a core,
but were by chance observed very early in their formation, and hence only momentarily
appear as \Sersic{} because they did not develop yet an measurable core.
Core-\emph{less} galaxies however include also ``pure'' \Sersic{} galaxies, which
will never develop a core because their formation e.g.\ involved significant gas
fractions --- these are contaminant objects we want to segregate.
Lacking an age estimation, one indicative way to select ``pre-destined'' core
galaxies is to adopt the magnitude limit: it is in fact observed that ETG brighter
than $M_{B} \sim -20.5$~mag are statistically prone to present cores
\citep[e.g.][and references therein]{kormendy:2009,graham:ETG_review}.
We therefore applied this magnitude cut to the core-\emph{less} \cite{richings}
galaxies with a measured $\Sigma$ (or $T_c$), resulting in the selection of NGC~0474 and
NGC~4125\footnote{
 We excluded NGC~5982 for coherency, being one of the few objects for which
 \citealt{richings} did not report a distance, although the distance
 retrieved from NED technically makes this object part of the selection.
}.
Such a small number of matching objects (2 out of the parent 41) was indeed expected since
observing a \corser{} galaxy soon after its formation is a statistically rare occurrence:
the vast majority core-\emph{less} objects are ``pure'' \Sersic{} galaxies.
These two objects appear in panel B-1, where their mass deficit has been set to an
arbitrarily small value.
Under the assumption that NGC~0474 and NGC~4152 are in-the-making core galaxies,
their location in the plot is consistent with our evolutionary scenario. 

Notice that the overall $\Sigma$ value does not distinguish between features
primarily associated with gas-rich interactions and star-formation (e.g.\ tidal
tails and streams) and those potentially associated with gas-poor mergers and/or no
star-formation (e.g.\ shells), despite its numerical value is more biased towards
shell structures \citep[see formulation in][]{schweizer:sigma}.
For our study, we are in principle interested in fine structures \emph{not}
associated with gas infall since this would hamper the formation of a depleted core
(see $\S$\ref{Introduction}).
To exclude the possibility that the fine structures of the core galaxies in our
sample were associated with significantly gas-rich mergers, we visually inspected
SDSS-DR13 images of the sample galaxies using the ``SDSS Navigator'' applet\footnote{
 \url{skyserver.sdss.org/dr13/en/tools/chart/navi.aspx}
}.
We found that the only fine structures presented by the galaxies were shells,
with the exception of NGC~4552 and NGC~5557, which also presented plumes.
This observation also suggests that the galaxies considered here underwent
similarly gas-abundant mergers, and in turn it supports the implicit assumption that
the significance of fine structure monotonically declines in time.

A similar argument can be constructed for the $T_c$ galaxies.
Differently form $\Sigma$, the tidal parameter is not particularly sensitive
towards a specific feature (although the name might suggest it is biased towards
tidal tails).
By inspecting the notes reported by \cite{tal} in their Table 1 (which describe
the appearance of the features) we identified only one galaxy, namely NGC~5061,
which could bear signs of a past wet merger.
\cite{tal} report that this object shows a pronounced tidal tail on top of shell
structures.
Notably, by virtually removing the contribution to $T_c$ due to the tidal
tail (i.e.\ solely considering the shells), the corresponding data point in
panel \emph{B-2} of Figure \ref{figure:M_ratio_Sigma_All} would fall even closer
to the best-fit line.  

\section{Discussion}
\label{Discussion}

\noindent
In the recent years, alternative formation scenarios for the excavation of a core
have been promoted.
The need for alternative models arose partly to explain the observation of a growing
number of objects with extremely large cores
\citep[$>$ 1~kpc; e.g.][]{hyde,postman,lopez-cruz,dullo:IC1101}
which would imply extremely large depleted masses.
In \cite{bonfini:A2261} we calculated that as much as $10^{11}$ \Msun{} have
been displaced from the center of the Brightest Cluster Galaxy (BCG) of Abell~2261,
and created a 3.6~kpc core, much more than what simulations predict for SMBH binary
scouring \citep[][]{merritt:2006}.

As a corollary to the binary SMBH scenario, it has been suggested that --- after
coalescing --- the merged SMBH will receive a kick due to the anisotropic emission
of gravitational waves: this might explain larger cores.
In fact, the displacement of the merged SMBH transfers large amounts of energy to the
surrounding stars, and this might not only directly generate a core
\citep[e.g.][]{merritt:recoil,boylan},
but also significantly enlarge an existing one, possibly by multiple oscillations
\citep[][]{gualandris:2008,lena}.
Among the emerging alternative scenarios, there is also the ``AGN feedback''
mechanism proposed by \cite{martizzi}, and the core excavation by an
in-falling, captured perturber \citep[e.g.][]{goerdt:perturber,petts:perturber}.
The latter was actually proposed to explain the formation of constant density
cores in dark matter halos, but in \cite{bonfini:A2261} we proved it could
also be applied to the study of extremely large baryonic cores.

However --- although few exceptional cases might require alternative models ---
the SMBH scenario is still the most favoured due to the extended theoretical
support, and the existence of scaling relations between the
core properties and the SMBH mass \citep[e.g.][]{lauer:2007a,dullo:2014}.
There are also growing observational indications for SMBH pairs at the kpc scale,
observed at the center of merger remnants (e.g. NGC~6240, \citealt{komossa};
Arp~299, \citealt{ballo}; 0402+379, \citealt{rodriguez}; Mrk~463,
\citealt{bianchi}).
The recent ``visual'' detection of a binary whose SMBHs are separated by a projected
distance of only $\sim$7~pc poses the strongest observational confirmation
so far \citep{bansal:compact_SMBHB}.
At the same time, we are aware that galaxies with the largest cores ($>$ 1~kpc)
also host the most massive SMBH in the Universe, plausibly produced by a coalesced
binary \citep[e.g.][]{laine,lauer:2007a}.
The recent, direct detection of gravitational waves in stellar-sized merging
black holes \citep[][\emph{LIGO} collaboration]{first_GW,GW151226,GW170104}
arguably constitutes the strongest evidence of black hole coalescence (even
if at a smaller mass scale).

Our results provide additional support to the SMBH scouring scenario.
The linear relation in the central panels of Figure \ref{figure:M_ratio_Sigma_All}
shows clear indication that the significance of fine structures scales inversely to
the relative depleted mass over 4 orders of magnitude in $M_{def}$/$M_{sph}$.
Remarkably, galaxies which will eventually develop a core (see NGC~4125 and
NGC~0474 in panel \emph{B-1}) consistently present indication of prominent fine
structure. 
This strongly suggests that galaxy relaxation and core excavation are indeed
happening in parallel.
Once the core is almost completely excavated and the SMBH binary coalesced (1~Gyr),
a galaxy stops moving diagonally along the relation and it starts moving vertically down
until the complete disappearance of fine structures (which could take up to 4~Gyr).
This evolutionary track is represented in the central panel of
Figure \ref{figure:M_ratio_Sigma_All} with grey arrows.
The observation that no galaxy with large $M_{def}$/$M_{sph}$ presents significant
fine structure, and that all featureless galaxies have large $M_{def}$/$M_{sph}$,
is coherent with this picture.

The observed trend it is instead at odds with stochastic core depletion
events, like the capture of an in-falling satellite \citep[e.g.][]{goerdt:perturber}.
On the contrary, a larger core would require a larger satellite infall, which in
turn would produce larger fine structures.
The considerations about the AGN-feedback scenario \citep[e.g.][]{martizzi}
are instead more complex.
In the first place, AGN activity acts on a way smaller temporal scale
\citep[$\sim$10$^{7}$;][]{kauffman:AGN_duty,hopkins:AGN_duty} than fine structure evolution.
Outflows are impulsive and might produce a core of arbitrary size at any stage
of the galaxy life.
Secondly, focusing on the objects in our analysis, we exclude that they involved
large gas fractions ($\S$\ref{Measurements of cores and fine structures}),
hence we do not expect a significant AGN feedback able to justify
the creation of the largest cores in our sample.
As a matter of fact, the \cite{martizzi} mechanism has been proposed for
BCGs, in which the large cool inflow of cluster gas can explain extreme accretion
events.
Finally, one could argue that large depleted mass would imply a strong AGN activity,
which in turn requires significant gas infall on the SMBH which is more prone to
happen when the gravitational potential is not relaxed yet, i.e.\ right after the
merger, when $\Sigma$ is still large (at odds with Figure \ref{figure:M_ratio_Sigma_All}). 
Furthermore, it has also been questioned whether merger events are associated with
a higher AGN activity in the first place \citep{li:AGN_mergers}.

\section{Conclusions}
\label{Conclusions}

We promote the idea that depleted stellar cores in ETGs are progressively excavated
in parallel with the disappearance of the fine structure features (remnant of the
last major interaction experienced by the galaxy).
This prediction is in line with the expectations from the hierarchical formation
scenario of galaxies, and with the SMBH mechanism for core depletion.

We observe that the timescale for core excavation and relaxation of the galaxy
potential are similar or in excess of 1 Gyr (from the merger event which assembled
ETG), hence large enough for the two processes to initially co-evolve.
We show observational evidence that the prominence of fine structure
anti-correlates with the amount of mass removed from the core
(normalized by the total mass of the host galaxy).
We argue that this anti-correlation shall hold at least within the first Gyr
from the assembly of the galaxy (i.e.\ until the core is fully developed), after
which time the fine structure evolves independently.
Our results support the SMBH scouring as the main channel (although not
necessarily the only one) for the formation of depleted stellar cores.

Deep follow up of ETGs known to host cores will allow to better constrain
the relation presented in Figure \ref{figure:M_ratio_Sigma_All}, and to reduce the
systematic uncertainties due to the parametrisation of the fine structure.
By properly quantifying fine structures, and by calibrating them against
numerical simulations, one we will be able to replace the fine
structure measurement (the y-axis) with a time proxy.
By interpreting Figure \ref{figure:M_ratio_Sigma_All} in this way, our approach
will therefore represent a sophisticated method to exploit the structure
of galaxies to estimate the age of a merged ETG, overcoming the known measurement
limitations related to their uniformly old stellar populations.


\section*{Acknowledgements}

\noindent
The authors wish to thank the anonymous referee for the detailed comments which
significantly helped to strengthen our conclusions. 
TB would like to acknowledge support from the CONACyT Research Fellowships.
OGM wishes to acknowledge support by the UNAM PAPIIT grant (IA100516 PAPIIT/UNAM).



\bibliography{bibliography}

\label{lastpage}

\end{document}